\RequirePackage{fix-cm}
\documentclass[natbib,smallextended]{svjour3}       
\smartqed  
\usepackage{graphicx}
\usepackage{color}
 \journalname{my journal}
\begin{document}

\title{ Solar Sector Structure
}


\author{Hugh Hudson        \and
       Leif Svalgaard \and
       Iain Hannah
}

\authorrunning{Hudson, Svalgaard, \& Hannah} 

\institute{H.~S. Hudson \at
              Space Sciences Laboratory \\
              University of California \\
              Berkeley CA 94720 USA
              Tel.: +1-510-643-0333\\
              \email{hhudson@ssl.berkeley.edu}           
           \and
           L. Svalgaard \at
              W.W. Hansen Experimental Physics Laboratory, Stanford University, Palo Alto CA USA
              \and
              I.~G. Hannah \at
              SUPA School of Physics \& Astronomy
              University of Glasgow
              Glasgow, UK 
}

\date{Received: date / Accepted: date}

\maketitle

\begin{abstract}:
The interplanetary magnetic field near has a characteristic ``sector'' structure that reflects its polarity
relative to the solar direction.
Typically we observe large-scale coherence in these directions, with two or four ``away'' or ``towards'' sectors per solar rotation, 
from any platform in deep space and near the ecliptic plane.
In a simple picture, this morphology simply reflects the idea that the sources of the interplanetary field lie mainly in or near the Sun, 
and that the solar-wind flow enforces a radial component in this field.
Although defined strictly via the interplanetary field near one~AU, recent evidence confirms that this pattern also appears clearly at the level of the photosphere, with signatures including not only the large-scale structures (e.g., the streamers) but also highly concentrated fields such as those found in sunspots and even solar flares.
This association with small-scale fields strengthens at the Hale sector boundary, defining the Hale boundary as the one for which the polarity switch matches that of the leading-to-following polarity alternation in the sunspots of a given hemisphere.

\keywords{First keyword \and Second keyword \and More}
\end{abstract}

\section{Introduction}
\label{sec:introduction}
The early years of human exploration of space beyond the Earth system saw the development of instrumentation capable of detecting the then-hypothetical solar wind, and then the magnetic field embedded within it.
Qualitative physical arguments had made it clear that the field  stretches out in the radial direction, and that it eventually must adopt a spiral pattern in the ecliptic plane \citep{1957Obs....77..109B,1957Tell....9...92A,1958ApJ...128..664P}.
The mean speed of the observed solar wind dictates that this spiral should have angle of about 45$^\circ$ to the radial at one AU, and that the essentially radial flow should take 4-5 days in transit.
Observations generally confirmed these rough ideas.

\begin{figure}
 \includegraphics[width=0.45\textwidth]{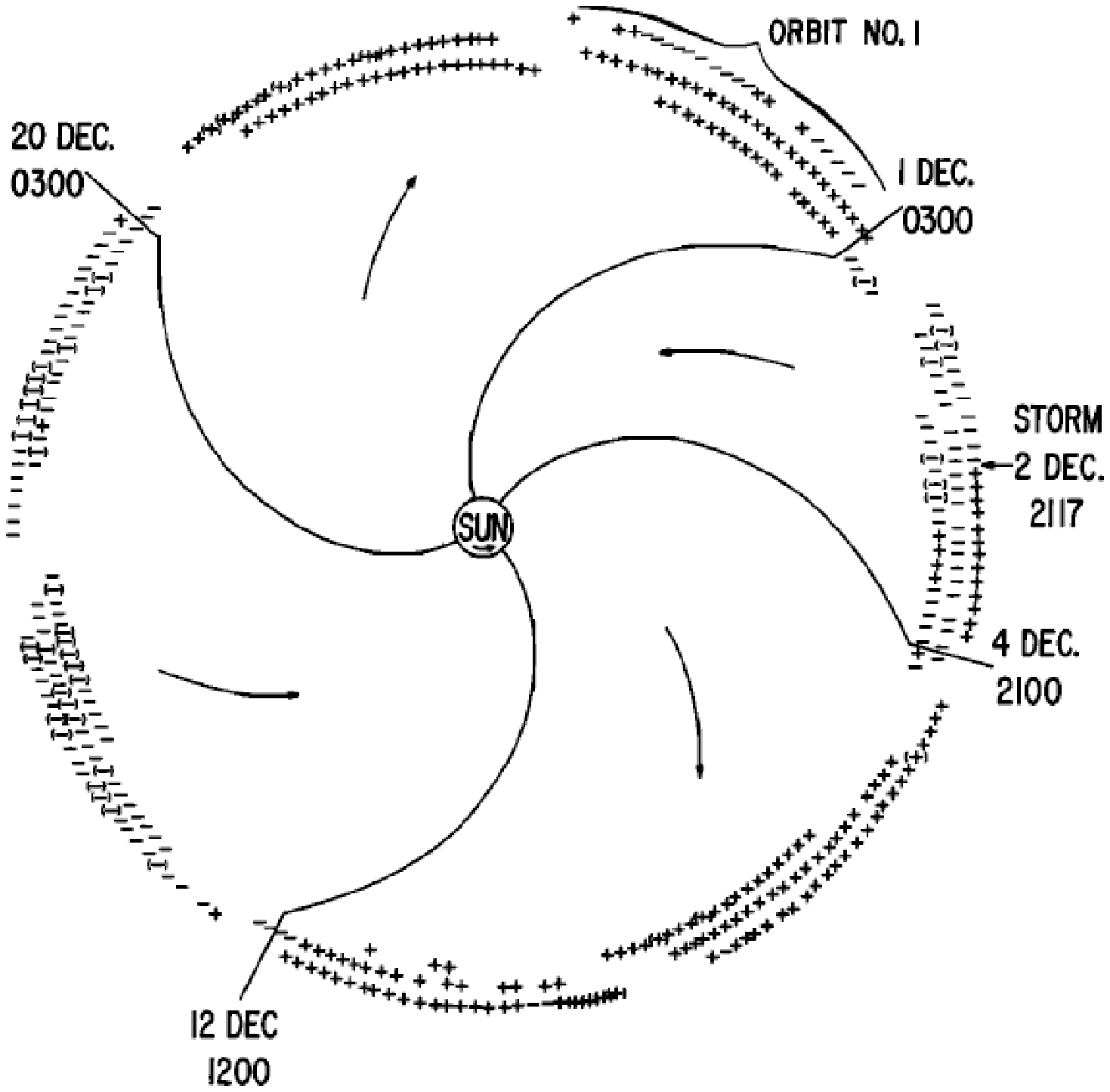}
\includegraphics[width=0.53\textwidth]{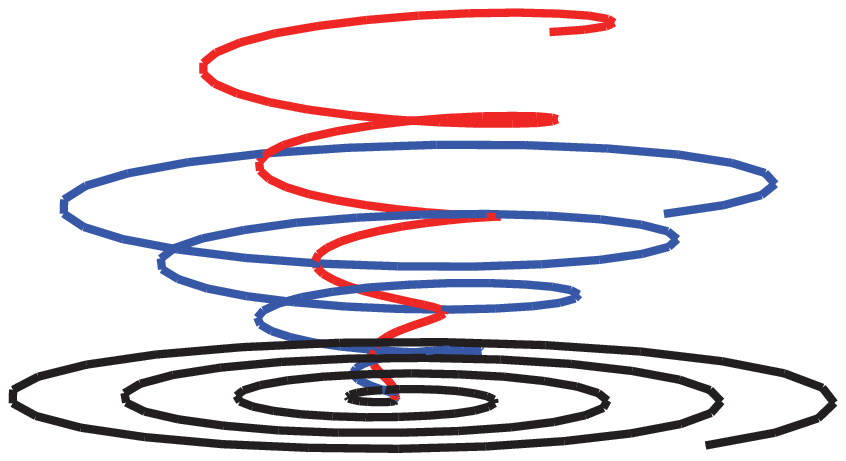}
\caption{Left, an early view of the interplanetary magnetic sector structure, from \cite{1965JGR....70.5793W}.
The symbols reflect the predominant orientation of the interplanetary field in 3-hour data segments from the magnetometer on board the IMP-1 spacecraft, anchored at Earth but with an apogee (31.7~$R_E$, orbital period about 3.9~d) large enough to give clear views of the solar wind in the ecliptic plane.
Plus and minus conventionally represent outward and inward field directions; this early sample shows a persistent four-sector pattern.
Right, a sketch of how field lines at 0, 30, and 60$^\circ$ heliolatitude must look in an isotropic high-beta solar wind with constant radial speed \citep[from][]{2013LRSP...10....5O}.
}
\label{fig:sectors} 
\end{figure}

The first interplanetary space probes capable of sufficiently sensitive magnetic measurements revealed the existence of the sector structure \citep{1965JGR....70.5793W}.
Figure~\ref{fig:sectors} (left) shows a data representation from that era, which illustrates the basic idea.
This remarkable feature of the solar magnetic field, as evidenced deep in the heliosphere but at heliolatitudes near the ecliptic plane, immediately demanded an understanding of the relationship between the different domains that the ``towards'' and ``away'' sectors represented.
The nature of the underlying solar magnetic field as revealed by magnetographic observations of the photosphere became a hot topic.
The 3D nature of the heliosphere (Figure~\ref{fig:sectors}, right ) compounded the complexity of these questions,  since the geometry of a sector could not be readily inferred from near-ecliptic observations.

In a major development, \cite{1972JGR....77.4027S} established that one can use polar geomagnetic records to determine the times when Earth passed through a sector boundary.
This opened the possibility of a proxy record of the global development of the heliospheric field that in principle can extend into the middle of the 19th century.
We discusss this in more detail in Section~\ref{sec:proxy} here.
Shortly thereafter Gulbrandsen (1973) and \cite{1974SoPh...36..115A} established a correlatopm between the sectpr pattern at one AU with coronal green-line (Fe {\sc xiv}) structures (Section~\ref{sec:pfss}) in the middle corona.
This finding directly related the heliospheric sector structure to small-scale solar magnetic fields.

In this review we describe (Section~\ref{sec:photosphere}) the recent further identification of the solar footprints of the sector patterns to the level of the photosphere (Svalgaard et al. 2011; cf. Svalgaard \& Scherrer 2014).
\nocite{2011ApJ...733...49S}
This involves the concept of the ``Hale sector boundary,'' defined by \cite{1976SoPh...49..177S} as that part of the sector boundary at which the polarity change matches that of the change of polarity from preceding to following sunspot (Section~\ref{sec:photosphere}).

In retrospect, it may seem reasonable that a heliospheric magnetic domain structure must exist, and that the domains must simplify into sector patterns at great distances from the solar surface.
The Sun possesses strong magnetic fields that originate in current systems largely contained below the photosphere; we know this because of the fairly good typical match with simple potential-field expansions in the lower corona (Section~\ref{sec:pfss}).
This means that if one makes a multipole harmonic extrapolation of the very complicated pattern of photospheric magnetic fields -- which evolves with large-scale organization, but also in seemingly random ways -- one predicts that the lowest-order term, the dipole, will eventually dominate the field and become the only surviving component at large radial distances.
This reasonable viewpoint ignores the fact that the field finds itself embedded in an active plasma, through which large-scale currents flow even far from the Sun.
In fact, the heliospheric magnetic field -- as opposed to the coronal field -- has a maximally non-potential structure according to the Aly-Sturrock theorem \citep{1991ApJ...375L..61A,1991ApJ...380..655S}.
It contains the large-scale current sheet needed to reverse the field direction between hemispheres.

\section{Solar magnetography and the sector structure}
\label{sec:wso}

Hale's discovery of Zeeman splitting in solar spectra has led to extensive, but tantalizingly incomplete, knowledge of the patterns of solar magnetism in the photosphere, and now to major efforts in the observational characterization of the magnetic field in the upper solar atmosphere; we return to this in Section~\ref{sec:pfss}.
In the era of discovery of the interplanetary sector structure, our remote-sensing knowledge of the solar magnetic field mainly consisted of synoptic maps, at low resolution, of the line-of-sight component.
Measurement of the full vector field has now become routine, but problems of interpretation remain; \cite{2011LRSP....8....4B} discuss many of them in a recent review.
We return to this topic in Section~\ref{sec:pfss}, and continue roughly historically here.

One essential problem is that resolution has improved to the point where the atmospheric dynamics and the characterization of the magnetic field require a simultaneous solution via numerical simulations.
The numerical approach will eventually help to solve the fundamental uncertainty in the physical heights of atmospheric structures. 
The standard semi-empirical modeling of the solar atmosphere \citep[e.g., the VAL-C model of][]{1981ApJS...45..635V} scales with optical depth,
rather than height directly, and has an upper boundary at 2543~km above the photosphere as normally defined in terms of optical depth unity at 5000~\AA.
Such models make many strong physical assumptions, for example in imagining a static structure that can be described as one-dimensional.
The density structure of the corona is best known only from the ill-posed 3D reconstructions of its projected brightness from projected data in one or two planes \cite[e.g.,][]{2008ApJ...679..827A,2008ApJ...680.1477A}, or via rotational synthesis that has inevitable confusion with K-coronal time variations \citep{2010SoPh..265...19F}.
Any quantitative understanding of field gradients must of course have precise geometrical information.
Nevertheless even the line-of-sight Zeeman measurements provide a wealth of morphological detail across the entire face of the Sun; we understand a great deal even at this level of observational technique.

\begin{figure}
\includegraphics[width=\textwidth]{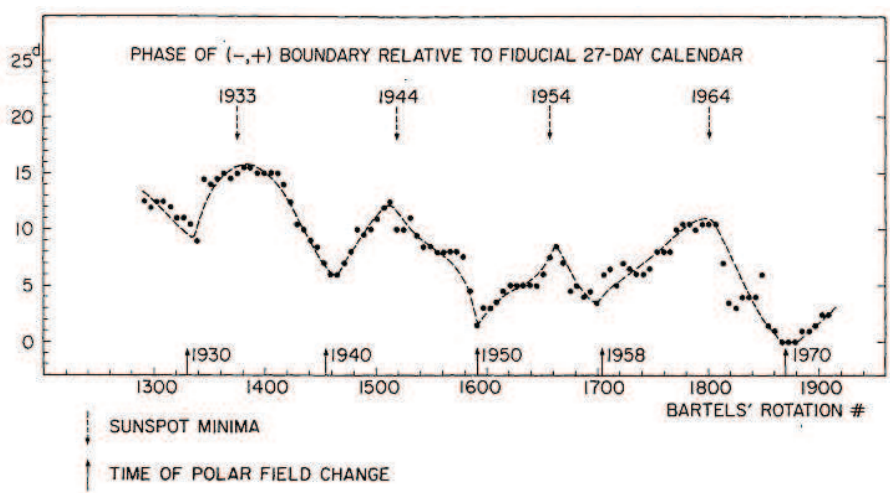}
\caption{Phase of sector-boundary crossing, relative to a fiducial 27-day rotation law \citep{1975SoPh...41..461S}.
}
\label{fig:sector_phase} 
\end{figure}

Typical line-of-sight solar magnetograms turned out to give no clues at all to the physical origin of the domains that the sectors represent, and early representations searched in vain for simple rigidly-rotating zonal patterns in the photosphere -- rigidly rotating, because one can derive a precise angular velocity from studying the phase of sector-boundary crossings.
Figure~\ref{fig:sector_phase} shows the stability of the sector pattern over a period of many years.
It represents the deviation of sector-boundary arrivals relative to a fiducial 27-day synodic rotation rate \citep{1975SoPh...41..461S}.
The stability of these phase measurements suggests a source that has only a slowly-varying rotation law.

Some of the mystery regarding the solar origins had been resolved by the discovery of the heliolatitude dependence of the sector patterns (the Rosenberg-Coleman effect), clearly recognizable in the early data from outside the ecliptic plane \citep{1969JGR....74.5611R}.
This led to the concept of a single warped current sheet \citep[e.g.][]{1973Ap&SS..24..371S,1975SoPh...45...83S} extending to large radial distances; its intersections with the ecliptic plane defines the sector boundaries seen there, and at sufficiently high heliolatitudes it must have a unipolar character; this concept extended the model proposed by \cite{1974SoPh...37..157S} and attained great prominence as the heliospheric ``ballerina skirt.'' 
The Ulysses mission triumphantly confirmed this picture \citep[e.g.,][]{2013SSRv..176..177B}, consistent with other data such as the radio (interplanetary scintillations) finding of high-speed solar wind streams at high heliographic latitudes \citep{1976JGR....81.4797C} and the X-ray finding of polar coronal holes \citep{1973SoPh...32...81V}.

The presence of an initially mysterious sector structure in the interplanetary field  \cite[e.g.,][]{1968SoPh....5..564W} led to the development of synoptic observing programs
(see also Section~\ref{sec:photosphere}).
Wilcox and his collaborators at Stanford University created what is now called the Wilcox Solar Observatory, which has provided a long-running data set of low resolution but great stability.
This facility began collecting data in 1975 and continues to the present time.

\section{Geomagnetism and the sector structure}
\label{sec:proxy}

The interplanetary magnetic field in the vicinity of the Earth interacts with the Earth's own field in a complicated manner; the orientation and strength of the field in the incoming solar wind perturb the geomagnetic field as measured on the surface of the Earth.
Given a terrestrial field of order 1~G and an interplanetary field of a only a few $\times 10^{-5}$~G, one might expect little effect, but in fact the signatures could readily be measured even with the techniques available in the 19th~century (the Gauss-Weber variometer).
The sector structure thus has a proxy record that extends quite far back in time.
\cite{svalgaard..1968} and \cite{1970Ge&Ae...9..622M} showed that the interplanetary magnetic field produced a characteristic pattern of diurnal variation at high-latitude sites such as Thule (12.5$^\circ$ geographic colatitude) or Vostok (168.5$^\circ$).
The Thule station has a geomagnetic colatitude of only about 5$^\circ$, and \cite{1973JGR....78.2064S} notes that the major effects of this perturbing current system lie with 15$^\circ$ of the geomagnetic pole; he attributes the characteristic variation to a Hall current \citep{1953RSPTA.246..281B} typically extending about that distance from the pole, and circulating with opposite senses for away and towards sectors.
The morphology of this diurnal effect distinguishes it sharply from others such as those of auroral electrojets.

\begin{figure}
\centering
 \includegraphics[width=0.6\textwidth]{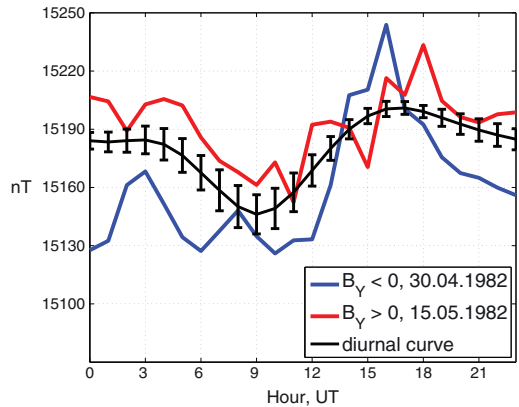}
\caption{Illustration of the Svalgaard-Mansurov effect using modern (1982) data from the Nurmij\"arvi station, which continued the 19th-century Helsinki observations.
The black line is the mean diurnal variations, and the red and blue are data from days of known away (red) and towards (blue) sectors, showing the clear distinction between the two orientations.
}
\label{fig:polar_currents} 
\end{figure}

This Svalgaard-Mansurov effect \citep{1972RvGSP..10.1003W} is strong enough to produce reliable determinations of the polarity of the interplanetary field \citep{1971NPhS..233...48F}, usually even on a daily basis \citep{2013GeoRL..40.3512V}, and so it serves to extend the record of Earth's sector-boundary crossing times back into the 19th~century.
This relies upon the understanding of the available geomagnetic observations at sufficiently high latitudes.
\cite{2013GeoRL..40.3512V} published sector-boundary data for Carrington cycles 9--13 (1843-1902) in this way, for example, by making use of the records from the now-discontinued Helsinki (29.8$^\circ$ geographic colatitude) and St. Petersburg (30.0$^\circ$) geomagnetic observatories.
Figure~\ref{fig:polar_currents} illustrates this heliospheric effect on the the Earth's polar field, calibrating the residuals away from the mean diurnal variation of Nurmij\"arvi data by reference to known sector orientations.
This observatory produces results closely matching those of the original Helsinki observatory \citep{nevanlinna97}.

\section{The corona, the solar wind, and the sector structure}
\label{sec:pfss}

In this section we discuss three perspectives on the sector structure: from the heliosphere, from the corona, and from the photosphere.
The information from these domains have different qualities (\textit{in situ} or remote-sensing), and are never complete, but generally agree now upon most of the basic structure.
We discuss one exception to this understanding in Section~\ref{sec:photosphere}.

\subsection{As viewed in the heliosphere}

\subsubsection{Observations {\rm in situ}}

The identification of solar sources of the interplanetary sector structure began with \cite{1973P&SS...21..703G} and \cite{1974SoPh...36..115A}, who found a remarkable long-term agreement between the sector boundaries, appropriately time-shifted, and coronal photometry in the ``green line'' of Fe~{\sc xiv} (5303~\AA).
We return to this important result in the next section, but first briefly describe the nature of the corona as seen in the green line or in various other ways.

One description of the solar corona envisions it as a concentric spherical shell of relatively low plasma beta (i.e., in force balance mainly via the Maxwell stress tensor); on intermediate time scales it is indeed quite stable but is prone to sudden and drastic disruptions, often on large scales, related to the domain structure of its magnetic connectivity.
The lower boundary -- subject to any of several possible definitions -- has corrugations imposed by the dynamics of the lower solar atmosphere, and the upper boundary fades into the solar wind in a manner that probably defies observation; both the upper and lower boundaries thus presumably have distinctly non-spherical shapes.

This complicated-sounding region has a simple and attractive model that describes its geometry fairly well: the ``potential-field source-surface'' (PFSS) model of \cite{1969SoPh....6..442S} and \cite{1969SoPh....9..131A}. 
Such models typically have a driver at the lower boundary from the line-of-sight Zeeman magnetic measurements, interpreted globally via a synoptic map of the measurements at central meridian.
Given the foreshortening toward the polar regions, this approach clearly omits a great deal of spatial and temporal structure that the EUV or X-ray movies, for example, reveal on many scales.
Based on this boundary, the PFSS model consists of a potential-field extrapolation out to an \textit{ad hoc} ``source surface,'' typically a spherical boundary at which a fictitious current system forces the exterior field into a strictly radial configuration.
This radial field then constitutes the ``open'' magnetic flux that somewhat mysteriously\footnote{How can a radial flow advect magnetic field parallel to itself?} fills the heliosphere via advection in the solar wind.
See for example \cite{2013LRSP...10....4L} for a discussion of subtle points in the estimation of heliospheric open field and its significance.
A common choice for the radius of the source surface, 2.5~R$_\odot$ \citep{1983JGR....88.9910H}, yields an approximate match to the heliospheric flux, which appears to vary by a small factor across the solar cycle.

\begin{figure}
 \includegraphics[width=0.41\textwidth]{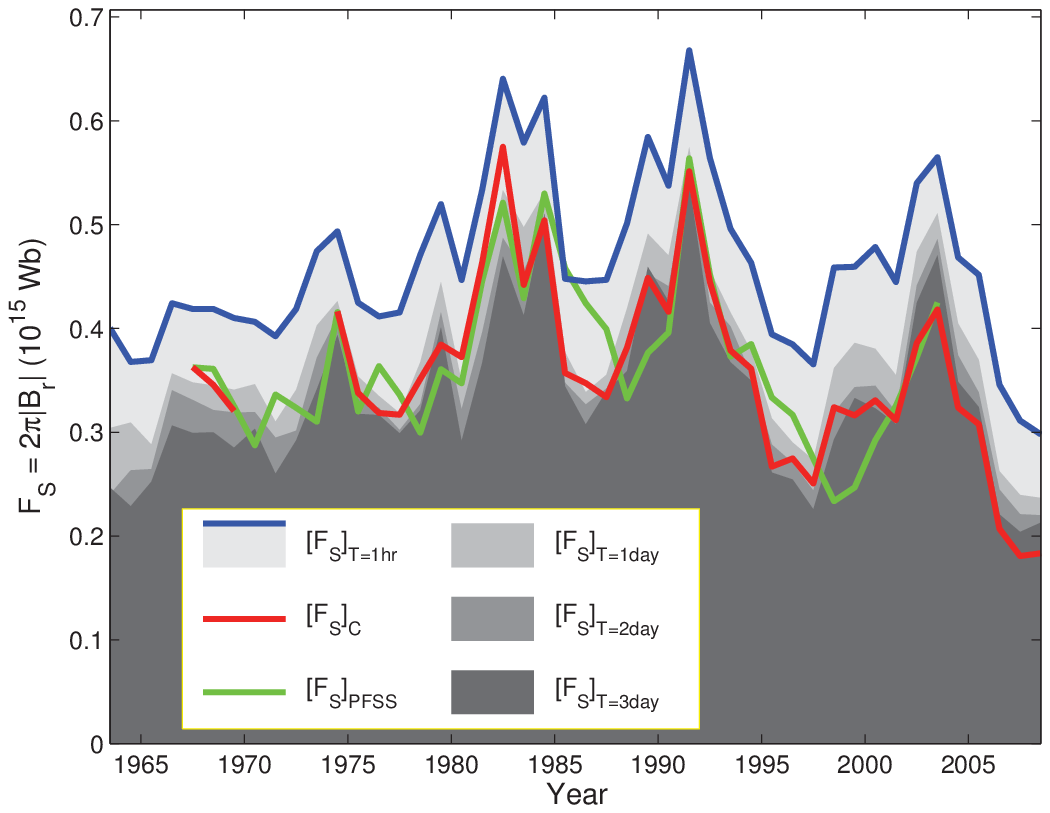}
 \includegraphics[width=0.57\textwidth]{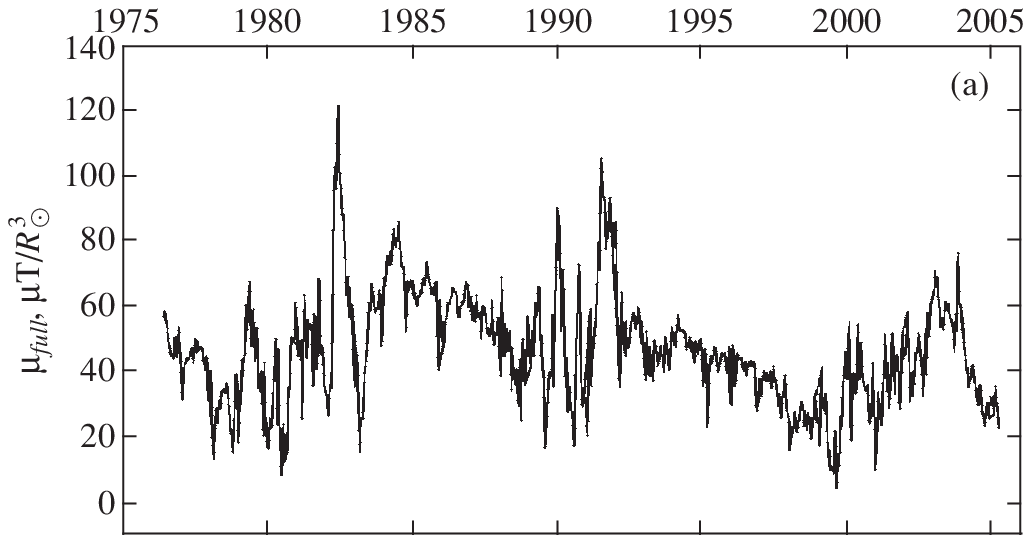}
\caption{Left, the solar open flux estimated via the PFSS method (red line) and heliospheric sampling (from Lockwood et al. 2009);
right, estimates of the amplitude of the best-fit dipole term for the photospheric field (Livshits \& Obridko, 2006).
}
\label{fig:lockwood_openflux} 
\end{figure}

We note that the heliospheric data make it possible to measure the open flux directly, in the sense of identifying it with the radial component  B$_r$ of the field.
The Ulysses data showed, remarkably, that there was a minimal latitude dependence of this quantity \citep{1995GeoRL..22.3317S}, so with that fact and the inverse square law, one can derive a good estimate of the open flux from any point in the heliosphere.
Again remarkably, given the weaknesses inherent in the photospheric data, the direct view in terms of B$_r$ turns out to agree reasonably well with that derived from PFSS with a fixed source surface \citep{2008JGRA..11312103O,2003ApJ...590.1111W}.
Figure~\ref{fig:lockwood_openflux} sketches the long-term variations of solar open flux as determined by these methods.

\nocite{2009JGRA..11411104L}
\nocite{2006ARep...50..926L}

\begin{figure}
 \includegraphics[width=\textwidth]{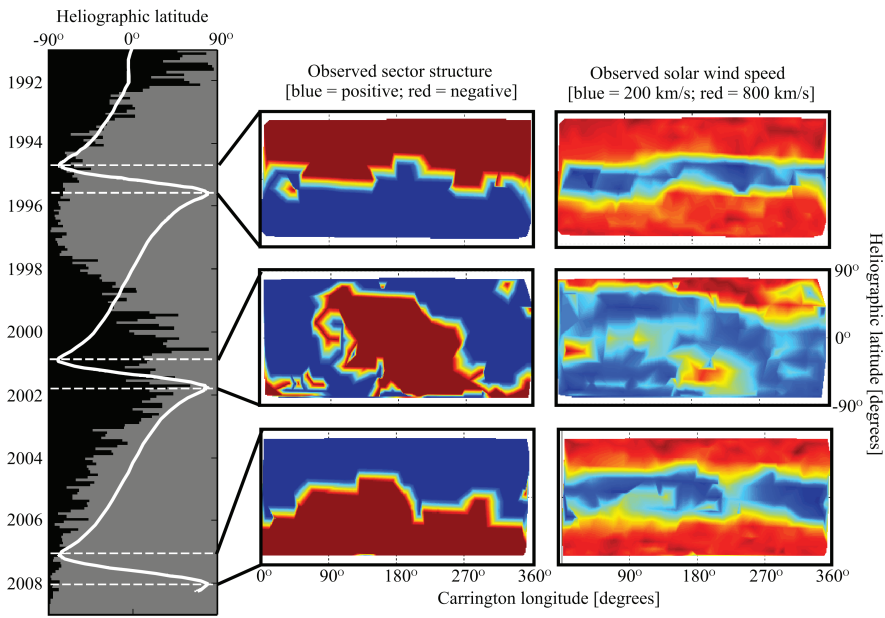}
\caption{The three rapid scans of heliolatitude executed by the Ulysses spacecraft: left, the heliolatitude and sunspot number; middle and right, source-surface projections of polarity and wind speed, respectively \citep[from][]{2013LRSP...10....5O}.
The Ulysses orbital inclination of 80.2$^\circ$ and perihelion of 1.3~AU meant that each full scan required about one year.
}
\label{fig:ulysses_overview} 
\end{figure}

The Ulysses ``fast latitude scans'' also provided an excellent overview of the 3D sector structure as it related to the solar wind, as illustrated in Figure~\ref{fig:ulysses_overview}.
This view nicely suggests how the basic pictures in Figure~\ref{fig:sectors} work themselves out in three dimensions, at least with this crude sampling; each map corresponds to the entire year of the ``fast'' scan and thus presupposes great long-term stability \citep{2011JGRA..116.4111O}.
Note that the small regions of minority polarity in the middle panels may just reveal artifacts resulting from this sampling; see \cite{2012SSRv..172..169A} for a theoretical discussion of the heliospheric magnetic connectivity.

\subsubsection{Remote-sensing observations}

\begin{figure}
 \includegraphics[width=\textwidth]{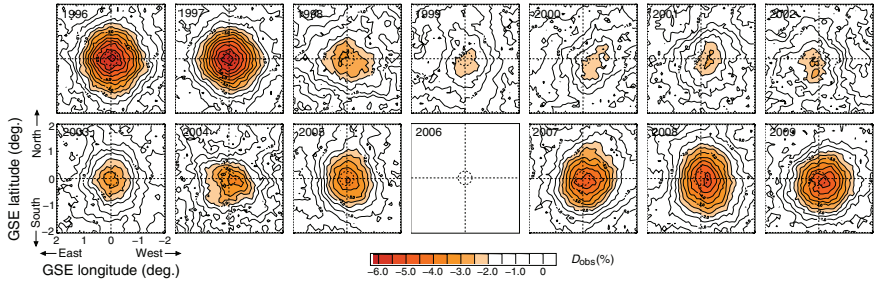}
\caption{The time development of the observed cosmic-ray shadow of the Sun, with its coronal magnetic field, from the Tibet air-shower array  \citep{PhysRevLett.111.011101}.
Year 2006 had insufficient statistics and the figure omits its image.
}
\label{fig:amenomori} 
\end{figure}

The idea of using the cosmic-ray shadows of large structures \citep{PhysRev.108.450} has developed into a method for probing the coronal magnetic field \citep{1993ApJ...415L.147A,PhysRevLett.111.011101}.
This observation uses TeV-energy primary cosmic rays detected via their extensive air showers, and the first results immediately showed a significant modification of the solar cosmic-ray shadow depending on the presence of towards and away sectors.
Since this 1993 paper, the data have accumulated and improved to the point where more sophisticated analyses for coronal magnetic-field structure have become possible.
Figure~\ref{fig:amenomori}, from \cite{PhysRevLett.111.011101}, shows the development of the cosmic-ray shadow detection over a solar-cycle time span.

A quantitative understanding of the solar magnetic shadow involves complex forward-method simulation.
Amenomori et al. estimate the FHWM angular resolution and modal energy of the Tibet air-shower data as 0.9$^\circ$ and 10~TeV, respectively.
Because of the small event rate, and the shallowness of the shadow, long integrations are required to obtain significant results.
These results depend upon Monte Carlo simulations of cosmic-ray transport and interaction within the inner heliosphere, for which a time-resolved magnetic model (e.g., the PFSS) is necessary.
The observations shown in Figure~\ref{fig:amenomori} permitted the authors to distinguish a standard PFSS model from a more elaborate current-sheet model \cite{1995AdSpR..16..181Z}.
More elaborate data analyses may allow us to follow the structure of the heliospheric field in the relatively unknown domain at the distance of the standard source surface, 2.5~R$_\odot$, and of course it would be most interesting to be able to characterize the development of the warp structure of the heliospheric current sheet in this region.

\subsection{As viewed in the corona}

Understanding the magnetic field in the solar corona directly requires remote-sensing astronomical measurements (e.g., for Zeeman splitting), which can be exceedingly difficult.
One can also interpret the images to obtain geometrical information from the orientations of the striations  \citep[e.g.,][]{2013ApJ...763..115A}.
The image-interpretation approach has improved continuously with the development of higher spatial resolution and the better sampling afforded by modern observations, because the images observable in the emission corona have striations at the finest observable scales.
These striations arguably guide us to the orientation of the field threading the plasma.
The gyroresonance condition also allows magnetography via microwave emission, especially in active regions; for example, \cite{2006ApJ...641L..69B}
found an example of kG fields at Mm heights via this technique.
None of these approaches though provides detailed direct information about {\bf B} everywhere in the corona.

Discerning the full geometry of the coronal field thus requires modeling.
The corona presents a serious geometrical problem here; for the most part it is optically thin, and so the line-of-sight depth cannot be known observationally.
The PFSS approach described in the previous section is the most frequently used method, but there are many other approaches with various degrees of sophistication \citep[e.g.,][]{2009ApJ...696.1780D,2012LRSP....9....6M}, mostly involving the assumption of a force-free field.
Beyond the purely mathematical treatments, which might suffice for an idealized corona at low plasma beta and no chromosphere, there are various efforts to incorporate plasma physics in various approximations, as also reviewed by Mackay \& Yeates.

\begin{figure}
 \includegraphics[width=0.8\textwidth]{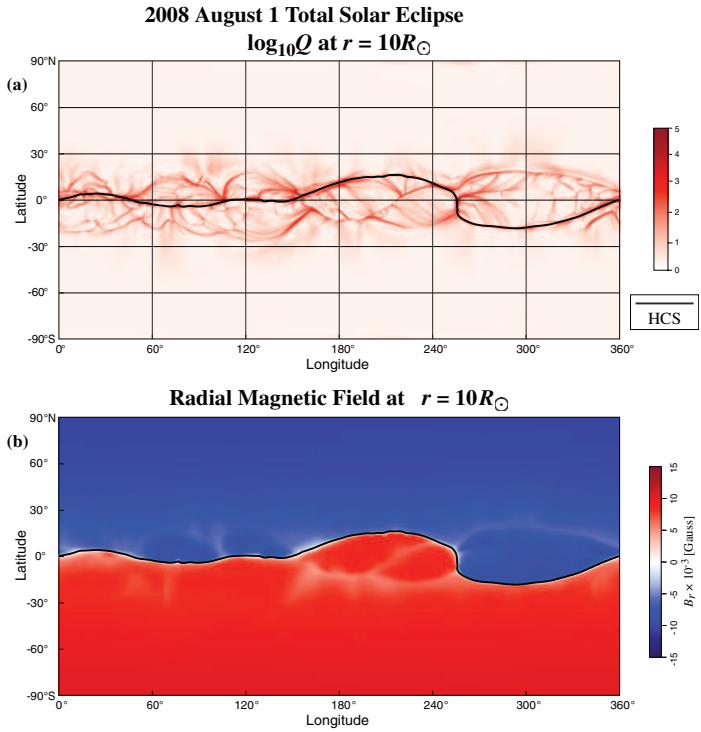}
\caption{Synoptic charts illustrating the ``Web of separatrices'' \citep{2011ApJ...731..112A} for a snapshot at solar minimum.
The upper panel shows the ``squashing factor'' Q, essentially the Jacobian of the mapping, with high values representing quasi-separatrix layers \citep{1995JGR...10023443P}.
The lower panel shows the radial field component B$_r$, and on both panels the black line shows the heliospheric current sheet.
}
\label{fig:antiochos_web} 
\end{figure}

The concept of the ``web of separatrices'' or S-web \citep{2007ApJ...671..936A,2011ApJ...731..112A,2012SSRv..172..169A} may simplify our view of how the complicated magnetic field of the lower solar corona maps into the relatively simple one in the heliosphere; in the ideal case of no sunspot activity whatsoever, this heliospheric field tends to become radial while retaining its bipolar character.
Figure~\ref{fig:antiochos_web} shows the S-web for a snapshot synoptic diagram at solar minimum.
Within the S-web structure one envisions magnetic reconnection to occur in such a manner that closed-field plasma, enriched in high-FIP elements, can escape into the slow solar wind over the its full breadth in heliographic latitude \citep[e.g.,][]{doi:10.1146/annurev.astro.45.010807.154030}.
A diffusive process with similar consequences underlies the ``interchange reconnection'' theories \citep[cf.][]{2001ApJ...560..425F,2009IAUS..257..109F}, but the details of both pictures remain somewhat unclear owing perhaps to the difficulty of modeling the local physics of magnetic reconnection within the MHD framework.

The synoptic chart in Figure~\ref{fig:antiochos_web} shows the interesting structure of the S-web: sometimes it appears to form mainly on one side of the heliospheric current sheet.
This azimuthal dependence could be related to the large-scale organization of the field embodied in the Hale boundary (Section~\ref{sec:photosphere}), but we are unaware of any work describing this possible association or its significance.

\subsection{As viewed at the photosphere}

A substantial development of solar magnetic-field models has proceeded entirely from their photospheric (surface) manifestations, the so-called flux-transport models following \cite{1964ApJ...140.1547L,1969ApJ...156....1L}.
Leighton's ``magneto\-kine\-matic model'' contains a few free parameters, including descriptions of a diffusive motion of vertical magnetic flux at the photosphere, and a meridional circulation there, and successfully describes the main properties of the solar cycle on this basis.
These models take their inspiration from the \cite{1961ApJ...133..572B} picture of how dynamo action may create the solar cycle deep in the interior of the Sun. Curiously, the Leighton flux-transport models succeed quite will in describing the surface manifestations without reference to the interior field structure at all.
The coronal field can be modeled in this way \citep[e.g.][]{1991ApJ...375..761W,2003ApJ...590.1111W}, and thence used to predict the properties of the heliospheric fields \citep{1983JGR....88.9910H}, including the locations of the sector boundaries.
These predictions do not require us to specify the interior volume threaded by the currents that give rise to the field outside the photosphere: deep-seated, as Babcock envisioned, or superficial, as modeled.

The interior-to-corona continuity of the large-scale solar field obviously has a descriptive representation in terms of spherical harmonics, and the behavior of the harmonic terms might give a guide to the physics involved.
As noted for example by \cite{GRL:GRL12399}, the lowest-order terms of the multipole expansion have the strongest effect at the source surface of a PFSS mode because it is at the outer boundary of the domain.
By construction, these terms determine the regions of open flux.
The variations of the axisymmetric and equatorial dipole terms therefore must dominate the open flux, as noted by many, if the main source of the interplanetary field lies within the body of the Sun.
Note the important caveat that we do not actually know how deep these sources may be (or what creates them), as opposed to the case of the Earth with its well-defined core.

\section{Photospheric identification of the sector boundaries}
\label{sec:photosphere}

The early observations of sector boundary crossings showed them to have stable phases relative to fixed periods (Figure~\ref{fig:sectors}).
This suggested a rigid rotation pattern, rather than a differential one, and yet no solar surface feature has such a property.
Subsequently the three-dimensional nature of the sector domains in the heliosphere became more apparent, as discussed above and illustrated in Figure~\ref{fig:ulysses_overview}.
At this point one could have inferred that the domain structure had a relatively simple interpretation: the waxing and waning of the polar coronal holes, and their pattern of polarity reversal, plus the simplification of the harmonic structure imposed by the expansion of the solar wind, could lead to a simplistic view of the structure formation.
The theoretical ideas discussed in Section~\ref{sec:pfss} confuse the issue by invoking differing views of the microphysics and its site.

\begin{figure}[htbp]
 \includegraphics[width=0.55\textwidth]{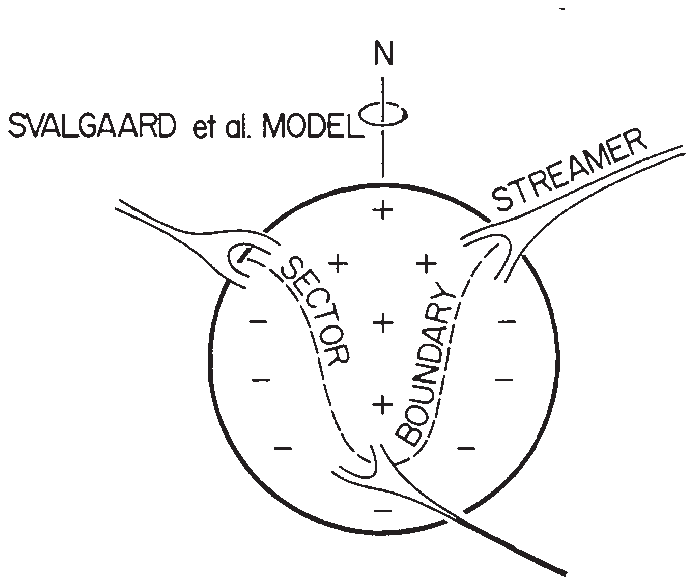}
 \includegraphics[width=0.43\textwidth]{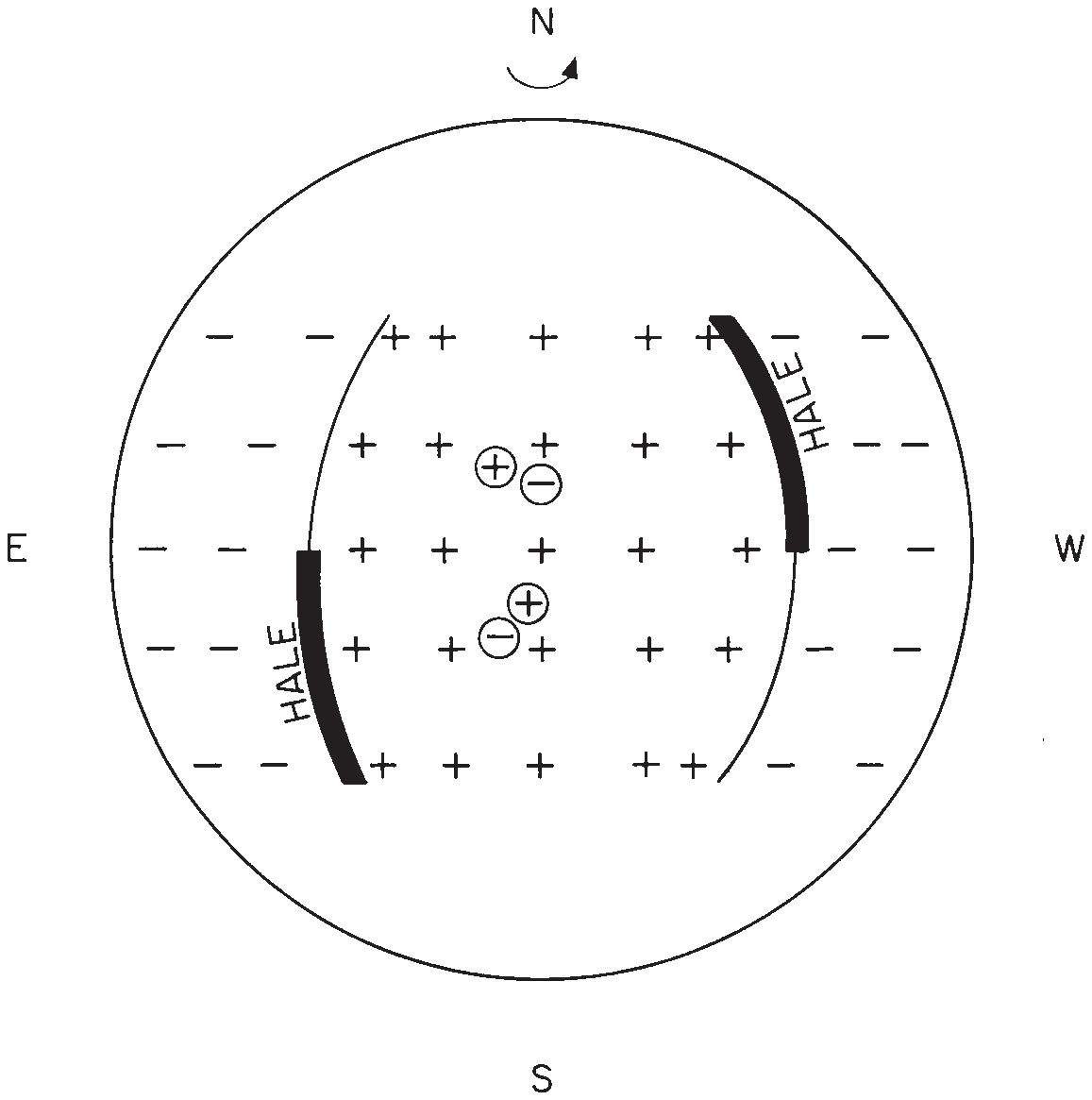}
\caption{Left, the now-accepted view of the coronal origin of the interplanetary sectors, as proposed originally by \cite{1974SoPh...37..157S}.
Right, the ``Hale boundary'' \citep{1974SoPh...36..115A}, defined as that part of the sector boundary at which the polarity switch matches that of the leading sunspot polarity in the corresponding hemisphere.
}
\label{fig:hale_boundary} 
\end{figure}

Recent work has opened new issues regarding the solar origins of the magnetic domains of the interplanetary sectors, extending the association first noted by \cite{1973P&SS...21..703G} via the green-line corona and its correlation with the sector boundaries.
The new work goes right to the level photosphere and identifies sector boundaries directly with small-scale features there, as we describe below. 
The key to these new associations is the concept of the Hale sector boundary \citep{1976SoPh...49..177S}, as illustrated in Figure~\ref{fig:hale_boundary}.
The recognition that the Hale sector boundary systematically correlated with signatures of solar activity could not have been anticipated, and still has no clear interpretation.
Geometrically, the sector structure itself seems fixed with respect to the solar rotation (see Figure~\ref{fig:sectors}); the various aspects of solar activity also do, though with latitude dependence reflecting the photospheric differential rotation.
So, what singles out the Hale boundary portion as a locus of solar activity?

Further and more specific evidence for the linkage of the Hale sector boundary with solar activity came with the discovery that solar flares preferentially occur at the Hale boundary \citep{2011ApJ...733...49S}, a result making use of the RHESSI flare catalog of flare locations (J.~McTiernan, private communication 2014).
This analysis reflects only the first few years of RHESSI observations, which began in 2002 and thus come from Cycle 23, but a further analysis (Iain Hannah, private communication 2013) extends this to Cycle 24 flares.
As noted by \cite{2011ApJ...733...49S}, RHESSI is not central here, and the ordinary NOAA flare listings also show the same effect but with reduced spatial resolution.

\begin{figure}[htbp]
 \includegraphics[width=\textwidth]{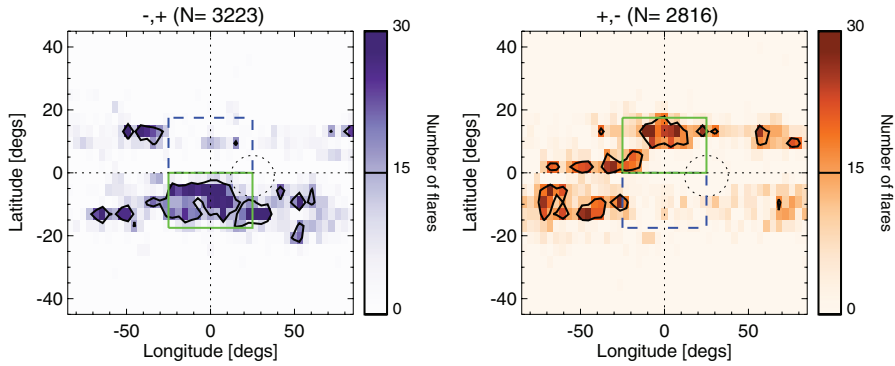}
\caption{RHESSI's view of the Hale boundary, for flare occurrence 2002-2010.
The colored contours show the locations of flares relative to the times of sector boundary crossings at one AU; the boxes with solid lines show the locus expected for the Hale boundary segment, and the dashed the disfavored. 
The left and right panels show the ($+,-$) and ($-,+$) crossings, respectively \citep[adapted from][]{2011ApJ...733...49S}.
}
\label{fig:rhessi} 
\end{figure}

\section{Long-term evolution of the sector structure}
\label{sec:evolution}

The long-term behavior of the solar magnetic field, extending many Hale cycles into the past, has theoretical as well as practical interest (the latter since some prediction techniques involve solar activity with patterns whose ``memory''  extends to the preceding cycle or even earlier). 
Regularities and irregularities exist on these longer scales, for example in the NS asymmetry discussed by \cite{1962SCoA....5..187B}.
This paper also made the link between sunspot asymmetry and an even stronger asymmetry in the occurrence of great magnetic storms over the solar cycles between 1833 and 1960.
A corresponding asymmetry appears in the Ulysses cosmic-ray data \citep{1996ApJ...465L..69S}, and
\cite{2003GeoRL..30.2135M} showed this cone-like distortion to be a persistent pattern of behavior.
\cite{2011ApJ...736..136W} then described this in terms of the additional magnetic pressure created in the low corona by the asymmetric eruption of field.

\begin{figure}[htbp]
\centering
\includegraphics[width=0.8\textwidth]{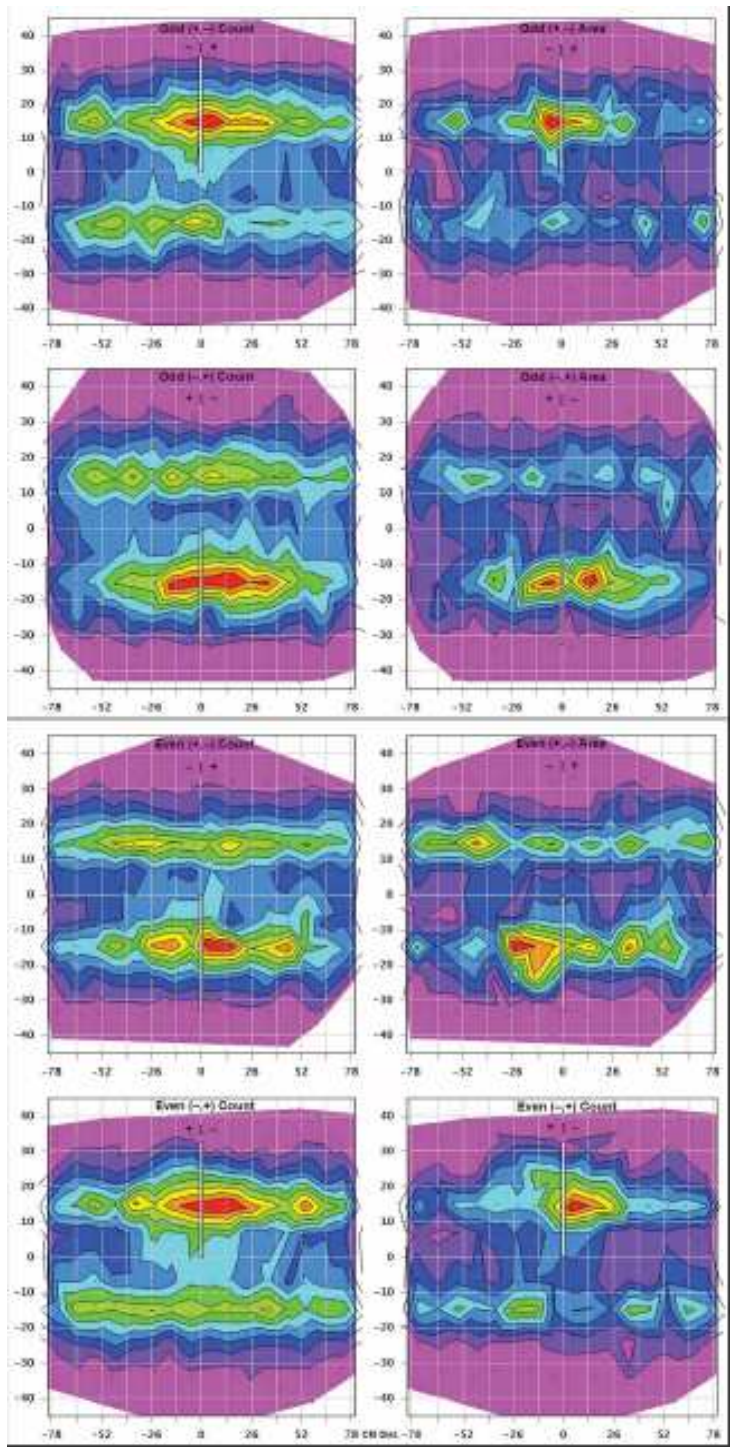}
\caption{Summed-epoch analysis for sunspot Group data, based on the Web archive maintained by David Hathaway, comprising some 600 sector boundaries identified for nine solar cycles (16 through 24).
The eight panels each show synoptic maps of relative longitude and latitude; the left column is for Group sunspot counts, and the right for areas.
The top four show cycles identified as ``odd'', and the bottom for the ``even'' ones, and in each case the Hale sector boundary is marked by a vertical rectangle.
}
\label{fig:Hale-Boundaries-and-Groups} 
\end{figure}

To the patterns of such long-term variations of sunspot, geomagnetic, and heliospheric variation we can now add the morphology of the Hale sector boundary, since it too can be traced well into the 19th century now.
Figure \ref{fig:Hale-Boundaries-and-Groups} shows the result of extending the historical record back to Cycle~9, in the form of summed-epoch analyses similar to those shown in Figure~\ref{fig:rhessi}: maps of occurrence summed on the key times given by the sector-boundary crossings identified geomagnetically, each showing the distribution as a function of central meridian distance and heliographic latitude.
The Figure shows four panels for each of the Odd cycles (16, 18, ..., 24) and Even cycles (17, 19, ..., 23), distinguishing the odd/even crossings from the even/odd crossings and in each case identifying the hemisphere of the Hale boundary. 
This confirms the association of sector boundaries with flux concentrations and the coincidence of Hale boundaries with the correct hemisphere in each case, and shows the result to be stable over multiple Hale cycles.

\section{What are the solar sources of the heliospheric field?}
\label{sec:sources}
An equivalent question might read ``Why do sector boundaries correlate with solar activity?''
The problem is that the small-scale closed fields of solar active regions, which lead through time to the structure of the streamer belt and the green-line observations \citep[e.g.,][]{1974SoPh...36..115A}, must appear as dome-shaped inclusions of more intense field in the low corona.
Conceptually, the dome becomes larger as more flux emerges, but then flattens out as the active-region magnetism disperses.
The streamer belt connects the active-region domes as illustrated in Figure~\ref{fig:hale_boundary}, left panel.
The solar wind forms above these magnetic domes or streamer belt; impelled by the large-scale sources of energy and momentum for the solar wind, the interplanetary field then develops its ballerina-skirt structure at higher altitudes.

The warp in the heliospheric current sheet has a natural explanation in terms of the intermittent dominance of activity in the N or S; roughly we could imagine that bigger magnetic domes or a more inflated streamer zone would simply push the heliospheric current sheet into the opposite hemisphere, considering only the structure resulting from magnetic pressure.
This however would put the active regions, sunspots, and flares in between the sector boundaries, not at their very location.
We do not have a ready explanation for this but suggest that it may have to do with the inherent latency of large-scale structure development in the corona.
\cite{2014SoPh..289..631Y} finds better fits to models with historical knowledge of the field development, rather than an instantaneous vacuum-field model.
In this case, the match between activity and sector boundary would be a coincidence, and a manifestation of the time taken for global coronal relaxation. 

\section{Conclusions}
\label{sec:conclusions}

The interplanetary magnetic sector structure offers opportunities and puzzles, as we have described.
One of the great opportunities lies in its proxy record, which when unfolded will give us a record of the largest-scale solar magnetic fields extending to at least Carrington cycle~9 (from about 1845).
The proxy extensions basically confirm that the Hale pattern of polar reversals persists over much longer time scales than the era of direct interplanetary observation, and these records may lead to further discovery.
For example, the Hale-boundary segregation extends back in time at least to Cycle 16 already, as described above.

As regards puzzles, we have the basic issue of the connectivity of the solar magnetic field between the photosphere and the heliosphere.
This broad subject involves several research communities, specifically those interested in the physical nature of the solar wind.
We require intricate 3D patterns of the sort represented in Figure~\ref{fig:antiochos_web}, and these must be dynamical rather than static.
Is there a link between the S-web (Antiochos) or interchange models (Fisk) and the Hale sector boundary?
The recent discovery of a strong association between the Hale boundaries and the small-scale fields of sunspots and their dynamics (flares) suggests that there might be.
The simplification of the photospheric fields as they map into the heliosphere has been found to be inherently dynamic, and so this connection seems like a reasonable one to suggest.

Another open puzzle has to do with the structure of the four-sector pattern. 
The literature may not convincingly describe the physics behind this persistent corrugation of the heliospheric current sheet (the higher-order wave of the ballerina's skirt).

\begin{acknowledgements}
We thank the International Space Science Institute for its hospitality during the preparation of this chapter.
Author Hudson thanks NASA for support under contract NAS 5-98033 for RHESSI.
\end{acknowledgements}

\newpage
\bibliographystyle{aps-nameyear}      
\bibliography{sectors}                

\end{document}